\documentclass[10pt,aps,prl,amsmath,amssymb,twocolumn,showpacs,byrevtex,superscriptaddress]{revtex4}
\usepackage{graphicx}

\begin{document}

\title{Is There a Higher-Order Mode Coupling Transition in Polymer Blends?}

%
%
\author{Angel J. Moreno}
\email[Corresponding author: ]{wabmosea@sq.ehu.es}
\affiliation{Donostia International Physics Center, Paseo Manuel de Lardizabal 4,
20018 San Sebasti\'{a}n, Spain.}

\author{Juan Colmenero}
\affiliation{Donostia International Physics Center, Paseo Manuel de Lardizabal 4,
20018 San Sebasti\'{a}n, Spain.}
\affiliation{Dpto. de F\'{\i}sica de Materiales, Universidad del Pa\'{\i}s Vasco (UPV/EHU),
Apdo. 1072, 20080 San Sebasti\'{a}n, Spain.}
\affiliation{Unidad de F\'{\i}sica de Materiales, Centro Mixto CSIC-UPV, 
Apdo. 1072, 20080 San Sebasti\'{a}n, Spain.}

\begin{abstract}
We present simulations on a binary blend of bead-spring polymer chains.
The introduction of monomer size disparity yields very different relaxation times
for each component of the blend. Competition between two different arrest mechanisms, 
namely bulk-like dynamics and confinement, leads to an anomalous relaxation
scenario for the fast component, characterized by sublinear time dependence 
for mean squared displacements, or logarithmic decay and convex-to-concave crossover 
for density-density correlators. These anomalous dynamic features, 
which are observed over time intervals extending up to four decades,
strongly resemble predictions of Mode Coupling Theory for nearby higher-order transitions.
Chain connectivity extends anomalous relaxation over a wide range of blend compositions. 
\end{abstract}
\date{\today}
\maketitle

\newpage

\begin{center}
{\bf I. INTRODUCTION}
\end{center}

The widespread potential applications of polymer blends have spurred
a great deal of recent research  on the dynamics of these systems showing
tunable properties. Dynamics in polymer blends are directly related with one fundamental
question of polymer-based soft matter systems: the way local friction arises
in complex environments that are chemically heterogeneous. Thermodynamically miscible polymer
blends exhibit complex dynamics. They show `dynamic heterogeneity' in the meaning
that two different segmental dynamics ($\alpha$-relaxation) can be observed in the blend.
Moreover, a symmetric broadening of any relaxation function of each component
is usually observed as the glass transition of the blend is approached. 
These features are usually understood in terms of 
self-concentration effects induced by chain connectivity \cite{lodge} and thermally 
driven concentration fluctuations \cite{kumar}.

For most of the investigated systems, the two components
in the blend display qualitatively similar dynamic features. 
However, recent experimental results by nuclear magnetic resonance \cite{lutz}
and neutron scattering \cite{genix} suggest that a rather different picture emerges
when the two blend components exhibit very different segmental mobilities in the
homopolymer state and the concentration of the fast component in the blend is rather dilute. 
Then, the segmental dynamics of the fast component strongly deviate from the
expected behavior and seem to be qualitatively different from that of the slow component.
Moreover, fully atomistic molecular dynamics simulations (MDS) suggest the possibility 
of an unusual logarithmic time decay for the intermediate scattering
function of the fast component \cite{genix}.

Simulations are a valuable tool for investigating the microscopic origin of these unusual features.
With present computational resources, a systematic study 
---i.e, by exploring a wide range of control parameters
as temperature or blend composition--- can only be performed by using very simple models,
providing they display the essential dynamic features observed in real systems. 
With these ideas in mind, in this article we present an extensive MDS investigation
on a blend of bead-spring polymer chains with components of very different mobilities. 
The fast component exhibits an anomalous relaxation scenario rather different from that
characteristic of usual liquid-glass transitions. Namely, mean squared displacements display
sublinear behavior, and density-density correlators exhibit a convex-to-concave crossover, which
is observed both by varying wavevectors and control parameters. By properly
tuning the latters, a logarithmic decay can be obtained. 

These anomalous features are observed
at intermediate time intervals extending up to four decades, and display striking
analogies with Mode Coupling Theory
(MCT) predictions for states close to a higher-order MCT transition, which emerges
as the result from the competition between different arrest mechanisms 
---we suggest bulk-like caging and confinement for the present case.

The article is organized as follows. In Section II we introduce the model and
give computational details. In Section III we present simulation results and evidence
that the fast component displays unusual relaxation features.
In Section IV the framework of the Mode Coupling Theory is used 
in an operational way to describe simulation results. Conclusions are given in Section V.

\begin{center}
{\bf II. MODEL AND SIMULATION DETAILS}
\end{center}

We have carried out simulations on a model very similar to one originally
introduced by Grest and Kremer \cite{grest} for homopolymers.
Here we investigate a binary mixture of bead-spring polymer chains.
Each chain consists of $N=10$ identical monomers of the species A or B.
Monomers of both species have equal mass $m=1$. The interaction between two given monomers, placed on
different chains or in a same one, is given by a soft-sphere potential plus a quadratic term:
\begin{equation}
V_{\alpha\beta} = 4\epsilon[(\sigma_{\alpha\beta}/r)^{12} - C_0 + C_2(r/\sigma_{\alpha\beta})^{2}],
\label{eq:potsoft}
\end{equation}
where $\epsilon=1$ and $\alpha$, $\beta$ $\in$ \{A, B\}.
The interaction is zero beyond a cutoff distance $c\sigma_{\alpha\beta}$, with $c = 1.15$.
The values $C_0 = 7c^{-12}$ and $C_2 = 6c^{-14}$ guarantee continuity of potential and forces at the
cutoff distance. The radii of the soft potential for the different types of interaction are
$\sigma_{\rm BB} =1$, $\sigma_{\rm AA} = \delta\sigma_{\rm BB}$, 
and $\sigma_{\rm AB}=(\sigma_{\rm AA}+\sigma_{\rm BB})/2$.
The potential (\ref{eq:potsoft}) is purely repulsive. It does not show local minima within 
the interaction range $r < c\sigma_{\alpha\beta}$. Hence, slow dynamics in the present model
arises as a consequence of steric effects, as in the computationally more involved case of
chains of hard-sphere monomers.

Adjacent monomers of a same chain also interact through a finite extensible
nonlinear elastic potential (FENE) \cite{grest,bennemann}: 
\begin{equation}
V^{\rm FENE}_{\alpha\alpha}(r) = -kR_0^2 \epsilon\ln[ 1-(R_0\sigma_{\alpha\alpha})^{-2}r^2 ],
\label{eq:potfene}
\end{equation}
where $k=15$ and $R_0 = 1.5$. The superposition of the potentials (\ref{eq:potsoft}) 
and (\ref{eq:potfene}) provides an effective bond potential for adjacent monomers with a sharp 
minimum at $r = 0.985\sigma_{\alpha\alpha}$, which makes 
bond crossing impossible for the investigated values of the monomer size disparity $\delta$.

The blend composition is defined as $x_{\rm B} = N_{\rm B}/(N_{\rm A} + N_{\rm B})$,
with $N_{\rm A}$ and $N_{\rm B}$ denoting respectively the number of A- and B-chains.
Data here presented for different compositions
correspond to a single value $\delta = 1.6$. Simulations were also carried out 
for values  $1 < \delta < 1.75$ at a single composition $x_{\rm B}=0.5$.
For $\delta \lesssim 1.2$ we did not observe signatures of the anomalous
relaxation features exposed below.

The packing fraction, $\phi$, is defined as $\phi = (\pi/6)L^{-3}[N_A\sigma_{AA}^3 + N_B\sigma_{BB}^3]$,
with $L$ the side of the simulation box. In the following, temperature, $T$, distance and time, $t$, 
will be given respectively in units of $\epsilon/k_B$, $\sigma_{BB}$ and $\sigma_{BB}(m/\epsilon)^{1/2}$.
Simulations have been carried out at a constant packing fraction $\phi = 0.53$.
This value is comparable to those used in simulations of slow relaxation in simple liquids \cite{notephi}.
 
We investigate the $T$-dependence of the dynamics for blend compositions 
in the range $0.1 \le x_B \le 1$. The total number of chains at each composition
varies between 250 and 600, with a minimum of 60 for the minoritary component. 
The system is prepared by placing the chains
randomly in the simulation box, with a constraint that avoids core overlapping.
Periodic boundary conditions are implemented.
Equations of motion are integrated by using the velocity Verlet scheme \cite{frenkel},
with a time step ranging from $2 \times 10^{-4}$ to $5 \times 10^{-3}$,
for respectively the highest and the lowest temperature. 
A link-cell method \cite{frenkel} is used for saving computational time
in the determination of the particles within the interaction range of a given one. 

At each state point, the system is thermalized at the requested temperature by periodic velocity rescaling.
After reaching equilibrium, energy, pressure, chain radii of gyration and end-to-end distances show no drift.
Likewise, dynamic correlators and mean squared displacements show no aging, i.e., no time shift 
when being evaluated for progressively longer time origins. 
Once the system is equilibrated, a microcanonical run is performed for production
of configurations, from which dynamic correlators and mean squared displacements 
are computed. For each state point, the latter quantities are averaged over
typically 20-40 independent samples.

\begin{center}
{\bf III. SIMULATION RESULTS}
\end{center}

\begin{center}
{\bf a) Mean squared displacements}
\end{center}

\begin{figure}
\includegraphics[width=0.90\linewidth]{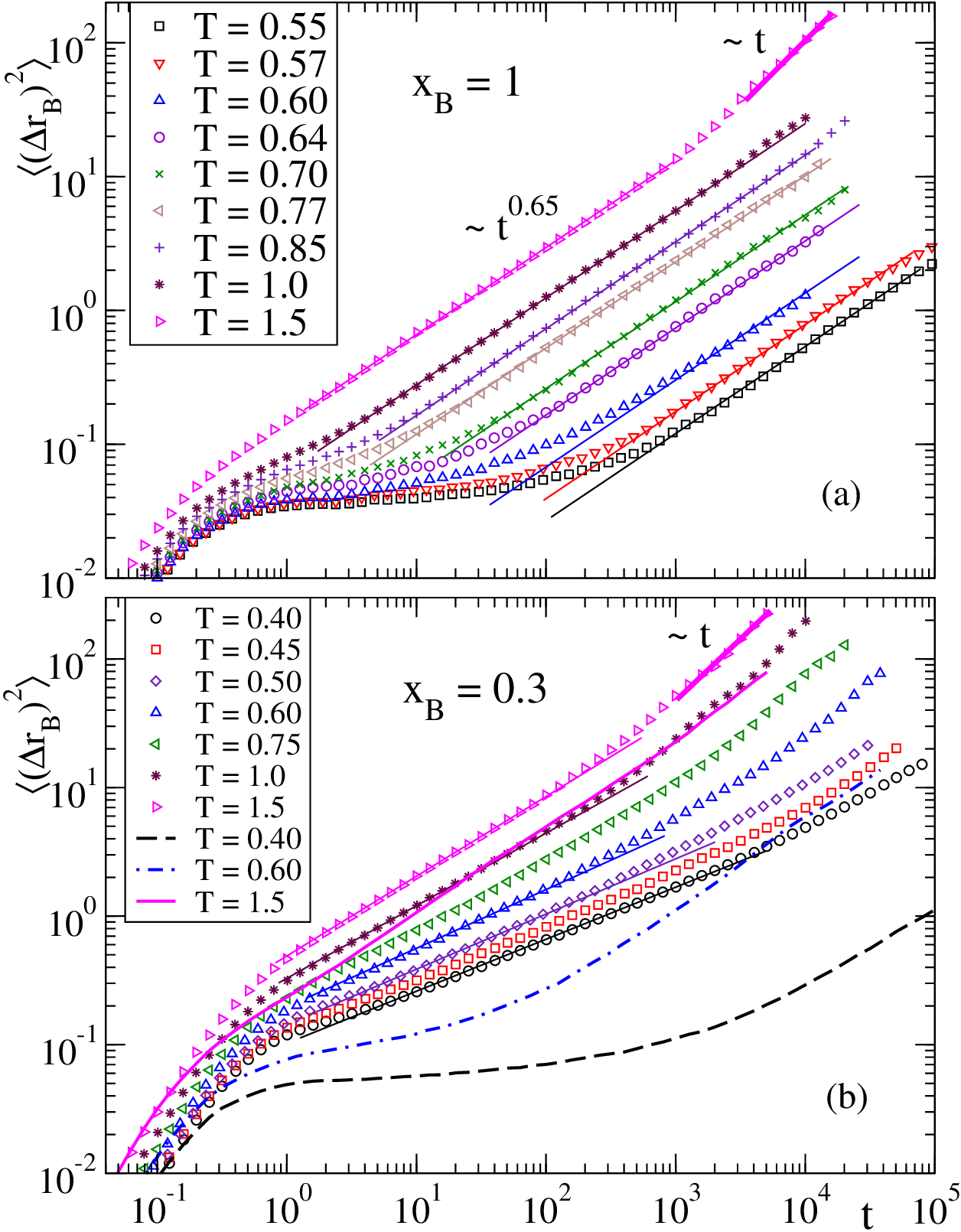}
\newline
\caption{Symbols: mean squared displacement of B-particles for the homopolymer (a) and
for a blend composition $x_{\rm B}=0.3$ (b). Thick straight lines correspond to the linear
diffusive regime $\langle (\Delta r_{\rm B})^2 \rangle \propto t$. Thin straight 
lines in panel (a) correspond to sublinear behavior $\langle (\Delta r_{\rm B})^2 \rangle \propto t^{0.65}$.
Thin straight lines in panel (b) correspond to 
$\langle (\Delta r_{\rm B})^2 \rangle \propto t^{\alpha}$ with a $T$-dependent exponent $\alpha$. 
From top to down, $\alpha=$ 0.635, 0.58, 0.47, 0.44 and 0.415. Results for A-particles (curves)
at $x_{\rm B}=0.3$ are also shown for comparison. The exponent for the 
observed intermediate sublinear regime is 0.66.} 
\label{fig1}
\end{figure}

Fig. \ref{fig1}a shows the $T$-dependence of the mean squared displacement of the B-particles,
$\langle (\Delta r_{\rm B})^2 \rangle$, for the homopolymer case ($x_{\rm B} = 1$).
We observe features analogous to similar bead-spring models previously investigated \cite{bennemann}. 
A plateau is developed after the initial ballistic regime ($t \lesssim 0.2$),
extending over longer times with decreasing temperature. 
This plateau corresponds to the well-known caging regime ---i.e., the temporary trapping 
of each particle induced by their neighbors--- 
observed when approaching a liquid-glass transition \cite{mctrev1,mctrev2}.
At longer times a crossover to a Rouse-like \cite{bennemann,aichele} sublinear regime 
$\langle (\Delta r_{\rm B})^2 \rangle \propto t^{0.65}$ occurs, before the final crossover
to linear diffusion at very long times, which is reached only for the highest investigated temperatures.
The $T$-independent exponent 0.65 is compatible with previous
simulations of similar models of bead-spring homopolymers \cite{bennemann}.

Fig. \ref{fig1}b shows results for A- and B-particles for a blend composition $x_{\rm B}=0.3$.
Monomer size disparity introduces a clear separation in the time scale
of both components. Hence, on average, the big A-particles and the small B-ones are respectively 
slow and fast particles. A-particles exhibit features
analogous to those observed in the homopolymer case.
It is noteworthy that this observation is fulfilled in the whole composition range,
suggesting that bulk-like dynamics is the only relevant arrest mechanism for the slow component.
However, in the range $x_{\rm B} \lesssim 0.8$, B-particles exhibit a rather different behavior, 
as illustrated in Fig. \ref{fig1}b for a composition $x_{\rm B} = 0.3$. 
The arrest mechanism is characterized by a larger 
localization length as compared to the homopolymer or to A-particles.
This is evidenced by the larger value of the mean squared displacement
at the inflection point after the end of the ballistic regime.
A direct crossover from the ballistic regime to sublinear behavior
$\langle (\Delta r_{\rm B})^2 \rangle \propto t^{\alpha}$ occurs, but differently from
the homopolymer or the A-particles, the exponent $\alpha$ is not constant, but shows
a clear decrease with decreasing $T$.  At sufficiently high temperature, again
$\alpha \sim 0.6$, ---i.e., above a certain $T$ there are no essential
differences between the transport mechanism for both species. 
However, for the lowest investigated temperatures we find $\alpha \sim 0.4$. 
%

\begin{center}
{\bf b) Density-density correlators}
\end{center}

\begin{figure}
\includegraphics[width=0.90\linewidth]{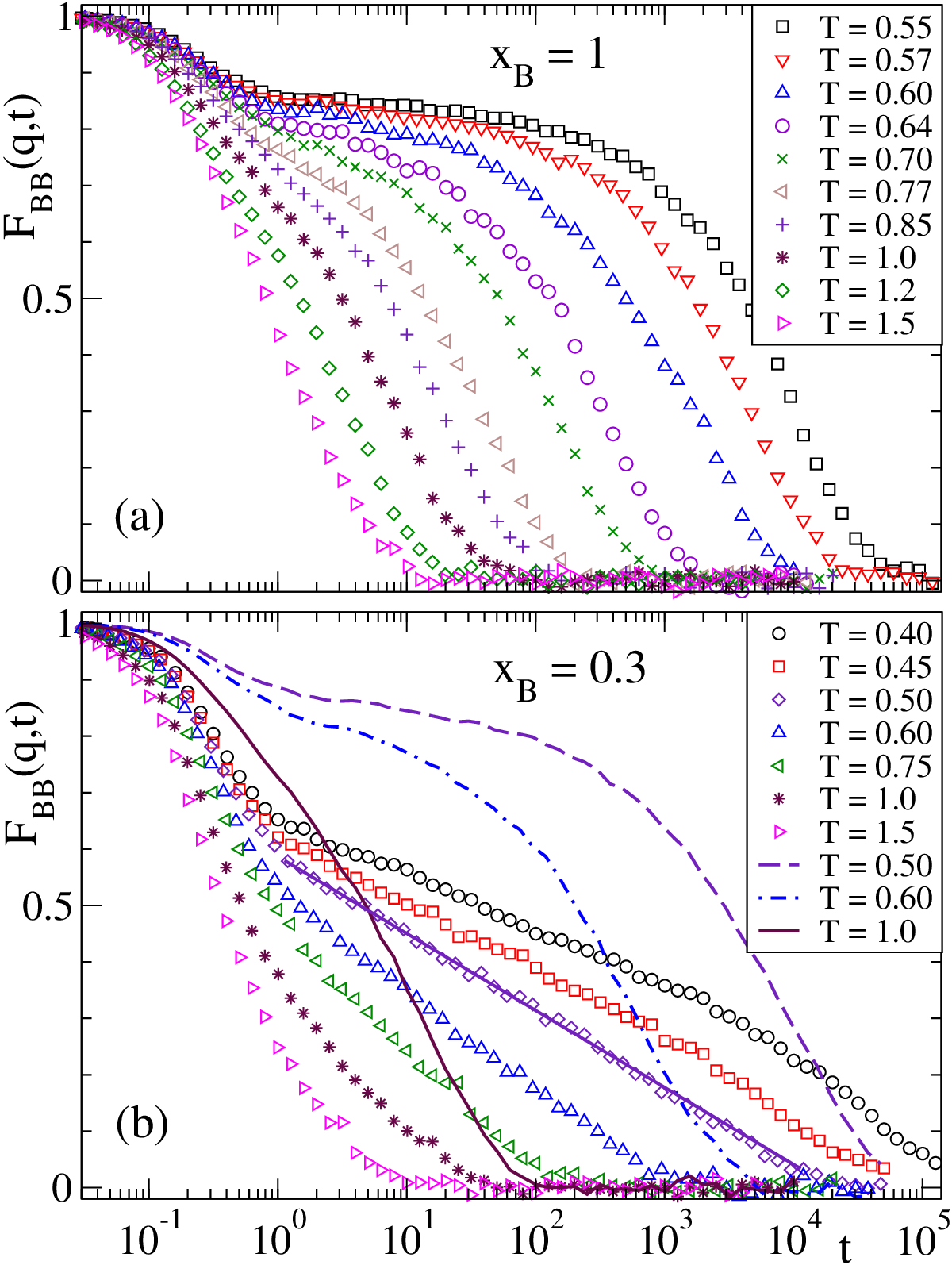}
\newline
\caption{Symbols in panels (a) and (b): $T$-dependence of $F_{\rm BB}(q,t)$, respectively
for the homopolymer and for a blend composition $x_{\rm B}=0.3$, as computed from the simulations.
The straight line in panel (b) indicates logarithmic decay over four time decades.
For $x_{\rm B}=0.3$, the correlator $F_{\rm AA}(q,t)$ is also shown (curves) for comparison. 
The wavevectors are $q=7.1$ for the homopolymer
and $q=4.5$ for $x_{\rm B}=0.3$, both values corresponding to the maxima
of the respective total static structure factors.} 
\label{fig2}
\end{figure}

\begin{figure}
\includegraphics[width=0.90\linewidth]{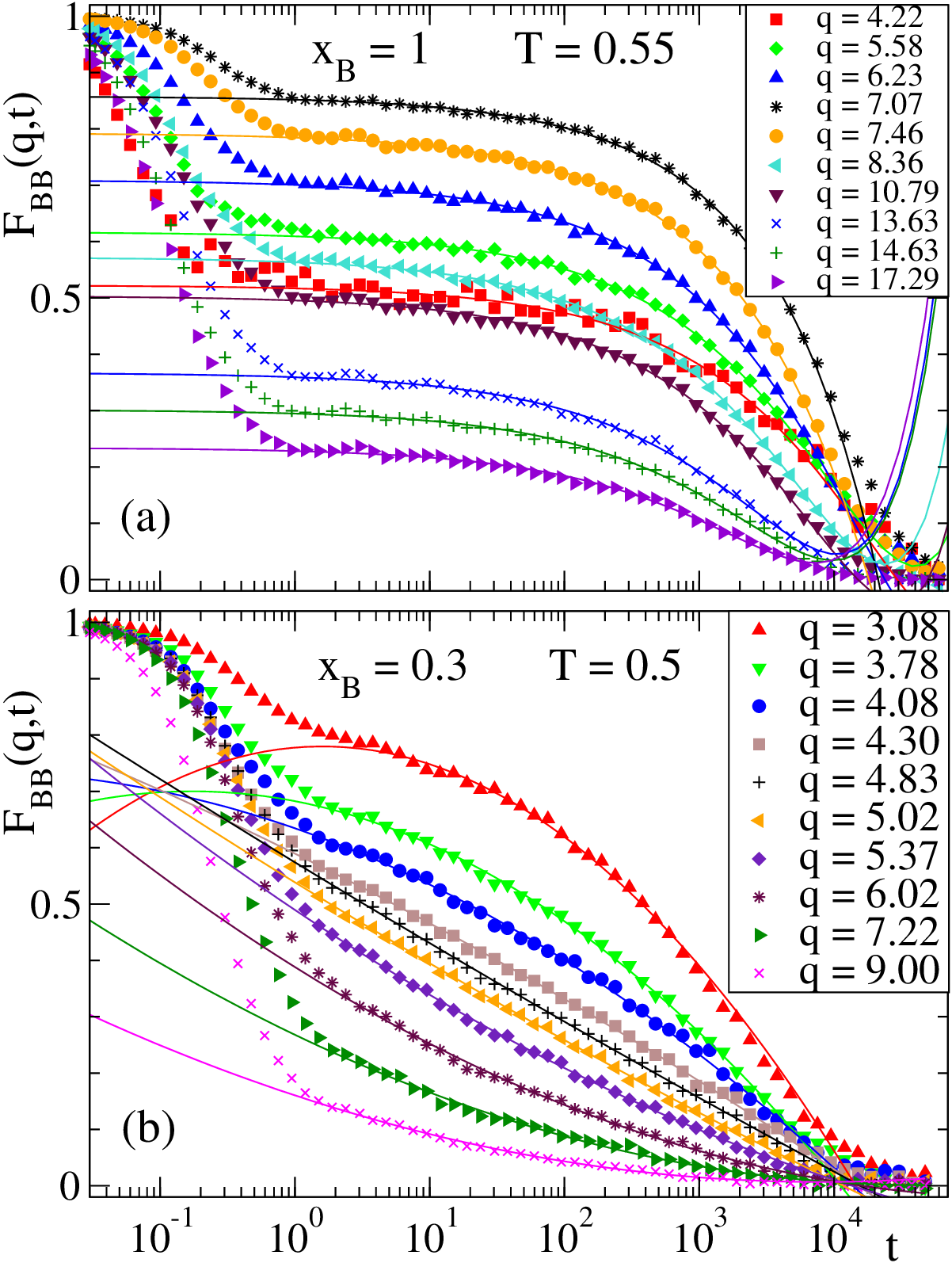}
\newline
\caption{Symbols in panels (a) and (b): $q$-dependence of $F_{\rm BB}(q,t)$, respectively for the 
homopolymer and for a blend composition $x_{\rm B} = 0.3$, as computed from the simulations 
at two selected low temperatures. Curves in panels (a) and (b) are respectively fits
to Eqs. (\ref{eqvonsch}) and (\ref{eqlog}).} 
\label{fig3}
\end{figure}

We compute partial density-density correlators 
$F_{\alpha\beta}(q,t)=\langle \rho_{\alpha}(q,t)\rho_{\beta}(-q,0)\rangle/
\langle \rho_{\alpha}(q,0)\rho_{\beta}(-q,0)\rangle$,
with $\rho_{\alpha}(q,t) = \Sigma_{j}\exp[i{\bf q}\cdot {\bf r}_{\alpha,j}(t)]$, 
the sum extending over all the particles of the species $\alpha \in$ \{A,B\}. 
Remarkable differences between the homopolymer and the blend are observed for density-density
correlators for B-B pairs, $F_{\rm BB}(q,t)$.  
Figs. \ref{fig2}a and  \ref{fig2}b show simulation results as a function of $T$ respectively 
for the homopolymer and for a blend composition $x_{\rm B}=0.3$.
The wavevector $q$ corresponds to the maximum of the total static structure factor.
Figs. \ref{fig3}a and \ref{fig3}b show the corresponding $q$-dependence at two selected low values of $T$.
The homopolymer exhibits the standard behavior in the proximity
of liquid-glass transitions \cite{mctrev1,mctrev2,das,bennemann,aichele}. After the initial transient regime,
$F_{\rm BB}(q,t)$ shows a first decay to a plateau. With decreasing temperature,
the plateau extends over longer time intervals. At long times, a second decay
occurs from the plateau to zero. This second decay corresponds
to the $\alpha$-process of the glass transition
and is well described by a stretched exponential function.
Analogous results are obtained for $F_{\rm AA}(q,t)$ in the 
whole composition range, as expected from the results above presented for the 
mean squared displacement of A-particles.

Rather different features are observed for $F_{\rm BB}(q,t)$ for compositions 
$x_{\rm B} \lesssim 0.8$. As can be seen in Fig. \ref{fig3}b, $F_{\rm BB}(q,t)$ 
does not exhibit a defined plateau at any value of $q$ or $T$.  
Moreover, an unusual convex-to-concave crossover, both in $q$ and $T$, 
is observed for the shape of the correlators, and in particular 
there are state point-dependent values of $q$ for
which logarithmic relaxation occurs over time intervals extending up to four decades.
Recent atomistic simulations \cite{genix}, 
at a single temperature and composition, of (fast) poly(ethylene oxide), PEO,
in a (slow) matrix of poly(methyl methacrylate) report an impressively similar crossover
at the $q$-dependence for self-correlations of PEO hydrogens (see Fig. 8 in Ref. 4), 
supporting the simple bead-spring model for polymer blends here investigated.

\begin{center}
{\bf IV. DISCUSSION}
\end{center}

The unusual features observed for mean squared displacements and dynamic correlators
of the fast component strongly resemble those reported by simulations \cite{zaccarelli,puertas,prlsimA4}
and experiments \cite{mallamace,pham,eckert} for hard-sphere colloids with short-ranged attractive interactions.
For these systems, such features have been recently rationalized in terms of the MCT
\cite{prlsimA4,fabbian,bergenholtz,dawson,sperl,jpcmA4}.
Motivated by this fact, we discuss the present results by using MCT in an operational way.
In its ideal version ---i.e., not including activated hopping events--- MCT predicts \cite{mctrev1,mctrev2,das}
a sharp transition from an ergodic liquid to a non-ergodic arrested state (`glass') at a given value
$x_{\rm c}$ of the relevant control parameter $x$ (in practice $\phi$ or $T$).
When crossing the transition point from the ergodic to the arrested state, the long-time limit
of the density-density correlator for wavevector $q$, $F(q,t)$,
jumps discontinuously from zero to a finite value, 
denoted as the critical non-ergodicity parameter, $f^{\rm c}_q$.
In the MCT formalism, the standard liquid-glass transition is of the {\it fold} type   
(also denoted as A$_2$) \cite{mctrev1,mctrev2,das,noteorder}.  The initial part of the $\alpha$-process 
(von Schweidler regime) is approximated by a power law expansion \cite{mctrev1,mctrev2,das}:
\begin{equation}
F(q,t) \approx f^{\rm c}_q -h_q (t/\tau)^{b} + h_q^{(2)}(t/\tau)^{2b},
\label{eqvonsch}
\end{equation}
where the prefactors $h_q$ and $h_q^{(2)}$ only depend on the wavevector $q$. 
The characteristic time $\tau$ only depends on the separation parameter $|x - x_{\rm c}|$
and diverges at the transition point. The exponent $b$ is 
related to the so-called exponent parameter $1/2 \le \lambda < 1$,
via the relation $\lambda=\Gamma^{2}(1+b)/\Gamma(1+2b)$,  
with $\Gamma$ the Gamma function \cite{mctrev1,mctrev2,das}.

\begin{figure}
\includegraphics[width=0.82\linewidth]{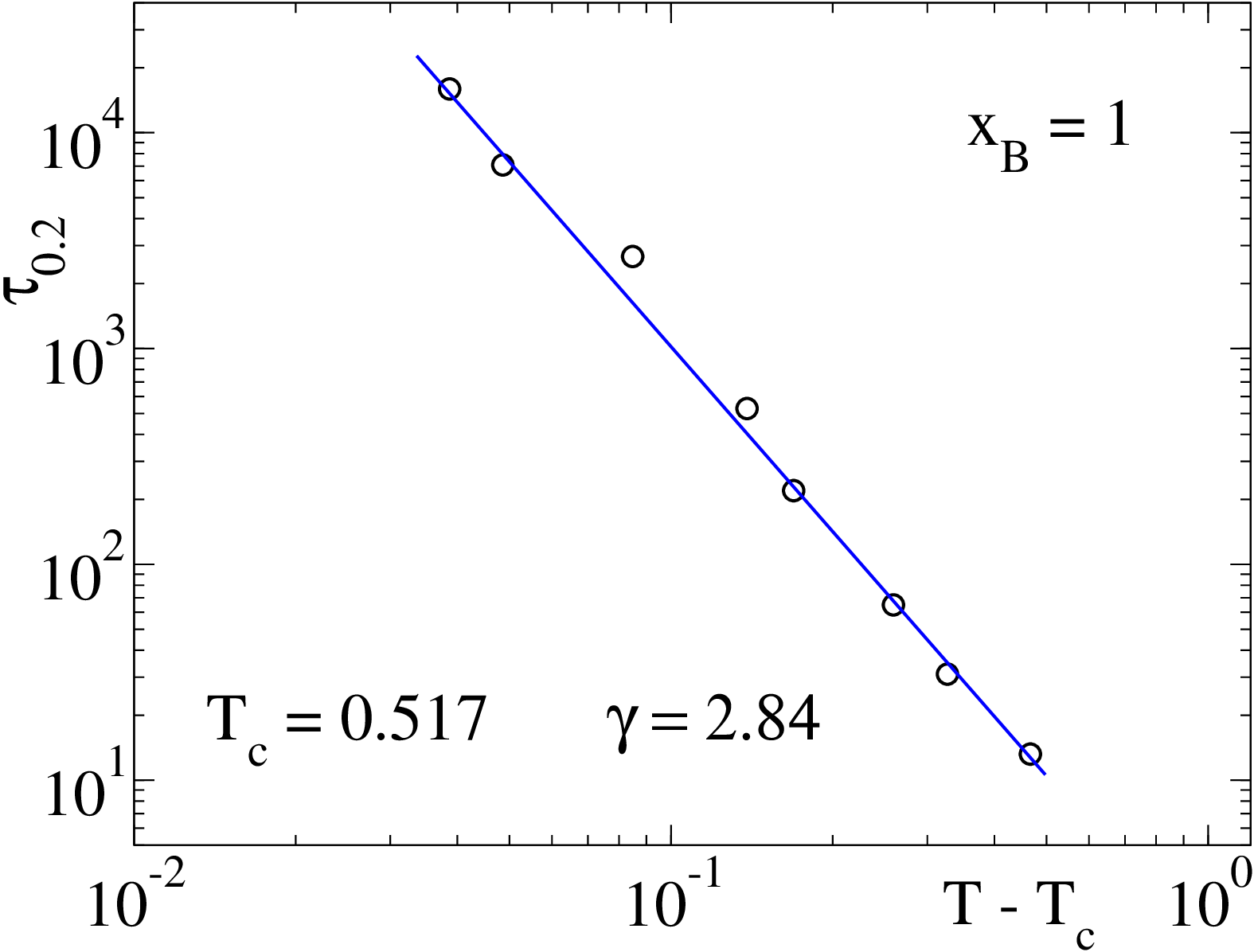}
\newline
\caption{For the homopolymer ($x_{\rm B}=1$), temperature dependence of the $\alpha$-relaxation 
time of $F_{\rm BB}(q,t)$ at $q$ corresponding to the maximum of the static structure factor.
Symbols are simulation data. The line is a fit to a power law 
$\tau_{0.2} \propto (T-T_{\rm c})^{-\gamma}$ (see text).} 
\label{fig4}
\end{figure}

As previously found for similar bead-spring models \cite{bennemann,aichele}, Eq. (\ref{eqvonsch})
provides a good description for the decay from the plateau in the homopolymer case
(see Fig. \ref{fig3}a). A value $b=0.477$ is obtained. 
The corresponding exponent parameter is $\lambda = 0.799$.
The value of the constant exponent $\alpha = 0.65$ for the observed 
Rouse-like regime in the mean squared displacement is consistent with MCT theoretical calculations
for chains of $N=10$ hard-sphere monomers ($\alpha = 0.6$) \cite{chong}.
The obtained $q$-modulation for $f^{\rm c}_q$ is also consistent with MCT expectations \cite{notehq}
and previous investigations of bead-spring models \cite{aichele}.

Fig. \ref{fig4} shows, for the homopolymer, the $T$-dependence of the relaxation time of $F_{\rm BB}(q,t)$,
for $q$ at the maximum of the static structure factor. We define the former as the time $\tau_{0.2}$ where
$F_{\rm BB}(q,t)$ decays to 0.2, a value well below the plateau. 
This corresponds to times probing the $\alpha$-relaxation. 
According to MCT for A$_2$-transitions \cite{mctrev1,mctrev2,das}, 
such times diverge at the transition point as a power law $\propto|x - x_{\rm c}|^{-\gamma}$ 
(in the present case $x=T$). The exponent $\gamma$ is related to $\lambda$ 
through the expressions $\gamma=1/2a +1/2b$ and $\lambda=\Gamma^{2}(1-a)/\Gamma(1-2a)$.
From the value $\lambda = 0.799$ given above we obtain $\gamma=2.84$. Consistently, a good description
of the data is obtained by forcing this value of $\gamma$ and leaving $T_{\rm c}$ as a free parameter
(see Fig. \ref{fig4}). We obtain $T_{\rm c}=0.517$.

The features observed for B-particles in the composition range $x_{\rm B}\lesssim 0.8$ cannot be rationalized
within the picture of a MCT A$_2$-transition. However, they exhibit a striking resemblance with
MCT predictions in the proximity of higher-order transitions A$_{n+1}$, characterized 
by $\lambda =1$, which can emerge as the result
from the interplay between $n \ge 2$ control parameters $x_1$, $x_2$ ... $x_n$.
Higher-order MCT transitions were initially derived for schematic models \cite{schematiczp,schematicpre}, 
and later for short-ranged attractive colloids \cite{fabbian,bergenholtz,dawson,sperl,jpcmA4}
as a first realization in real systems.
Close to a higher-order transition, or more generally to a fold transition with $\lambda \lesssim 1$,
an anomalous relaxation scenario emerges. The mean squared displacement exhibits an intermediate
sublinear regime \cite{sperl,notemsd} with a decreasing exponent as the transition is approached.
As reported above, this behavior is observed for $B$-particles for blend compositions $x_{\rm B}\lesssim 0.8$.
In the higher-order MCT scenario, $F(q,t)$ does not exhibit a defined plateau. Instead,
in an intermediate time interval of several decades, it is approximated by
a logarithmic expansion \cite{schematiczp,schematicpre,sperl,notefqc}: 
\begin{equation}
F(q,t) \approx f^{\rm c}_q -H_q \ln (t/\tau) + H_q^{(2)}\ln^2 (t/\tau),
\label{eqlog}
\end{equation}
where the prefactors $H_q$ and $H_q^{(2)}$ depend on $q$ and on the distance
of the state point $\{{\bf x}^n\}$ to the transition point $\{{\bf x}^n_c\}$.
As shown in Fig. \ref{fig3}b, Eq. (\ref{eqlog}) provides an excellent description
of the simulation data for $F_{\rm BB}(q,t)$ in the blend. It is worthy of remark
that the observed convex-to-concave crossover is present in
the higher-order MCT scenario \cite{schematiczp,schematicpre,sperl}. It is indeed one of its main signatures
and differentiates it from other theoretical frameworks \cite{sperl}.
The pure logarithmic decay observed at some $T$- and $q$-values is also present
in the higher-order MCT scenario, which predicts \cite{schematiczp,schematicpre,sperl} 
lines in the control parameter space with $H_q^{(2)}=0$. 



%

\begin{figure}
\includegraphics[width=0.90\linewidth]{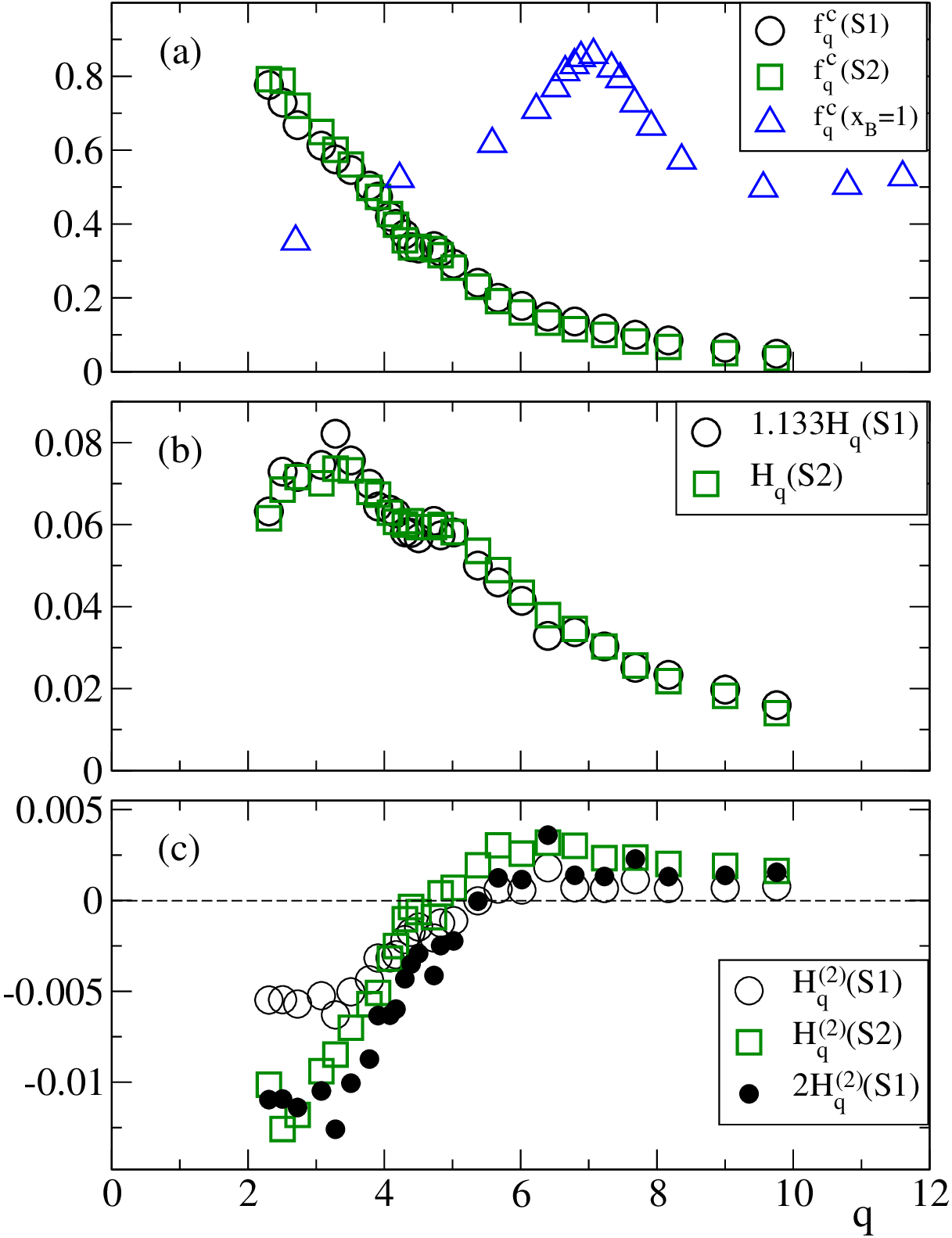}
\newline
\caption{Panels (a), (b), and (c) are respectively the values of $f^{\rm c}_{q}$, $H_{q}$, and $H_{q}^{(2)}$,
for two different state points S1 and S2,
as obtained from fits of $F_{\rm BB}(q,t)$ to Eq.~(\ref{eqlog}). The state points are
$S1$: ($x_{\rm B} = 0.3$, $T = 0.4$) and $S2$: ($x_{\rm B} = 0.3$; $T = 0.5$).
The characteristic times $\tau$ for S1 and S2 are respectively
1100 and 67. The $f^{\rm c}_q$ for the homopolymer ($x_{\rm B}=1$), 
as obtained from fitting to Eq.~(\ref{eqvonsch}), is also shown for comparison
in panel (a). The dashed line in panel (c) indicates zero value for $H_{q}^{(2)}$.}
\label{fig5}
\end{figure}

\begin{figure}
\includegraphics[width=0.90\linewidth]{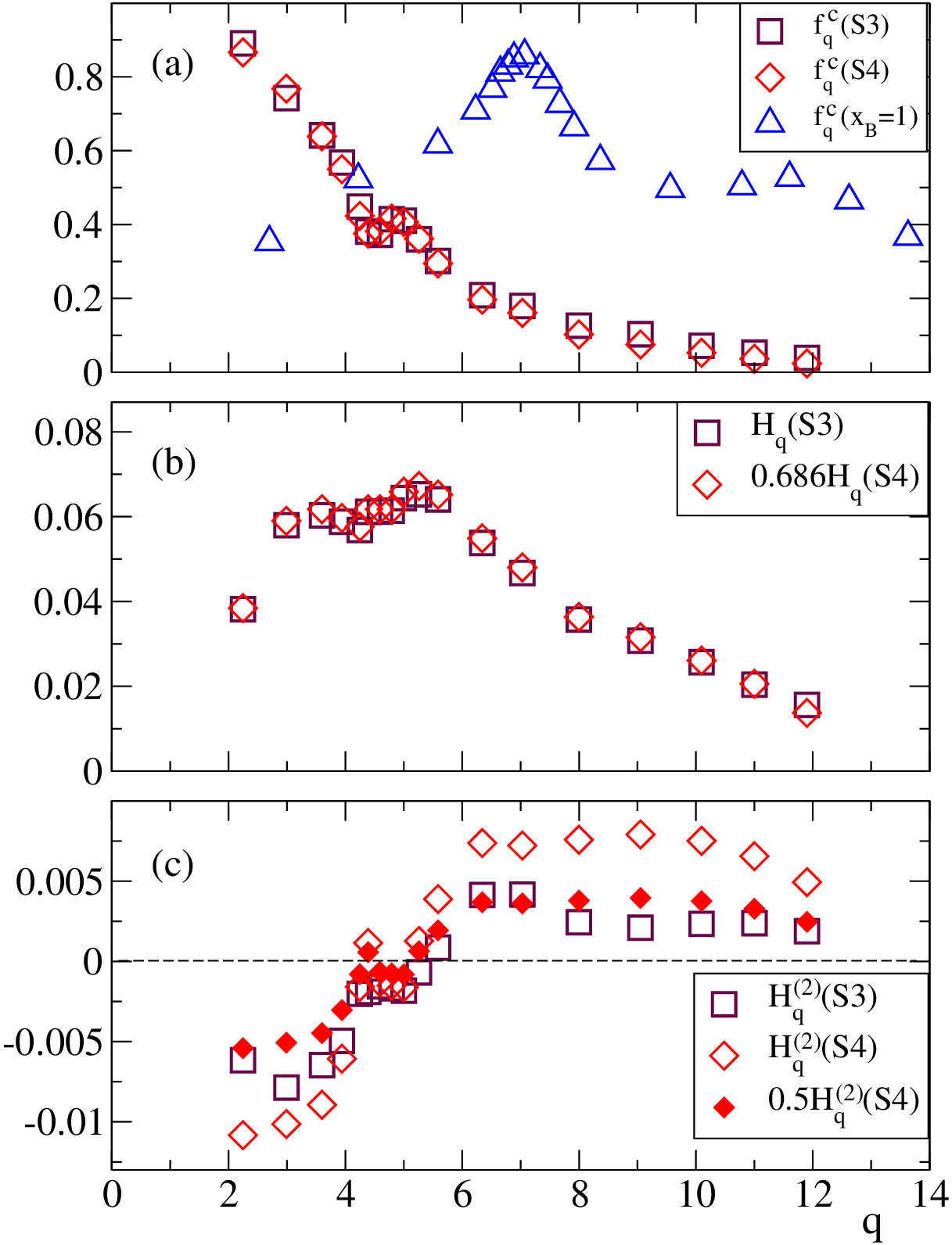}
\newline
\caption{As Fig. \ref{fig5} for the state points
$S3$: ($x_{\rm B} = 0.6$, $T = 0.4$) and $S4$: ($x_{\rm B} = 0.6$; $T = 0.6$).
The characteristic times $\tau$ for S3 and S4 are respectively
200 and 7.4.}
\label{fig6}
\end{figure}

Figs. \ref{fig5} and \ref{fig6} show the values of $f^{\rm c}_q$, $H_q$, and 
$H_q^{(2)}$ obtained from the fits to Eq. (\ref{eqlog}) of $F_{\rm BB}(q,t)$ for two sets
of state points [S1,S2] and [S3,S4]. The coordinates of such state points are
S1:($x_{\rm B}=0.3$, $T=0.4$), S2:($x_{\rm B}=0.3$, $T=0.5$),
S3:($x_{\rm B}=0.6$, $T=0.4$), and S4:($x_{\rm B}=0.6$, $T=0.6$).
The fact that a common value of $f^{\rm c}_q$ is found for the states of each set, 
together with the scaling behavior of $H_q$ (see below) would be consistent with the existence,
for each set, of a nearby MCT higher-order transition ($\lambda=1$) or 
a fold transition with $\lambda \lesssim 1$.  
Interestingly, $f^{\rm c}_q$ shows no strong modulation but a nearly monotonous decay 
(see Figs. \ref{fig5}a and \ref{fig6}a), 
resembling the qualitative behavior observed at A$_3$- and A$_4$-transitions
in short-ranged attractive colloids \cite{zaccarelli,prlsimA4,fabbian,sperl,jpcmA4}. 

According to MCT \cite{prlsimA4,sperl}, the coefficient $H_q$ in Eq. (\ref{eqlog}) factorizes as
$H_q = C(\{{\bf x}^n\}) \tilde{H}_q$, 
where $\tilde{H}_q$  only depends on $q$, and the $q$-independent 
term $C(\{{\bf x}^n\})$ depends on the state point.
Hence, the values of $H_q$ obtained for different state points close to the transition point
must be proportional. This scaling behavior is indeed displayed by simulation data, as demonstrated
in Figs. \ref{fig5}b and \ref{fig6}b. As for $f^{\rm c}_q$, the qualitative $q$-dependence of $H_q$ resembles
that observed at higher-order MCT transitions in short-ranged attractive colloids.

Also resembling MCT predictions for the higher-order scenario \cite{prlsimA4,sperl}, 
the coefficient $H_q^{(2)}$ is smaller than $H_q$, and does not obey scaling.
Indeed any test of scaling behavior over a wide $q$-range for $H_q^{(2)}$ is unsatisfactory 
(see two examples in Figs. \ref{fig5}c and \ref{fig6}c). Apparently, the value of $q$ for which $H_q^{(2)}=0$
is dependent on the state point, in agreement with MCT expectations.

The striking dynamic analogies of the bead-spring polymer blend here investigated
with the higher-order MCT scenario, 
which has been derived for short-ranged attractive colloids, provide connections with different research fields.  
In these latter systems, in a narrow range of $\phi$ and $T$, 
anomalous relaxation originates from the competition between two arrest mechanisms
of very different localization lengths: reversible bond formation
induced by the short-ranged attractive potential, and hard-sphere repulsion \cite{puertas,prlsimA4,pham,gelnmax}.
The dynamic analogies between so different systems suggest that this anomalous scenario 
might be a general feature of systems with several competing mechanisms for dynamic arrest,
and should motivate theoretical approaches within the MCT framework.
In the case here investigated, we suggest bulk-like caging and confinement
as the two competing mechanisms for arrest of the fast B-component. 
While the former is induced by the neighboring B-particles, the latter is induced by
the matrix formed by the chains of the slow A-component. 

It is worthy of remark  that, due to chain connectivity, 
any given particle has always neighboring particles of its same species.
Hence, competition between bulk-like caging and confinement for the dynamics of the fast component
occurs even at high dilution of the B-particles. Indeed, all the reported anomalous features 
are observed in a composition range from $x_{\rm B} \lesssim 0.8$
to, at least, the smallest investigated value $x_{\rm B}=0.1$.

Results reported here must not be understood as a proof of the existence of higher-order MCT transitions
(or fold transitions with exponent parameter $\lambda \lesssim 1$) in the system here investigated. 
A proper answer to this question could only be provided by solving the corresponding MCT equations.
Still, it is worth mentioning that recent theoretical calculations by Krakoviack \cite{krakoviack} 
have reported a higher-order MCT transition for a binary mixture of mobile and static hard spheres,
which can be seen as a simplified model for liquids in confining media with interconnected voids.
This system, where competition between bulk-like dynamics and confinement occurs,
is {\it a priori} closer to polymer blends than short-ranged attractive colloids. This result
supports the hypothesis of an underlying higher-order MCT scenario in the model here investigated.

\begin{center}
{\bf V. CONCLUSIONS}
\end{center}

We have carried out simulations on a bead-spring model for polymer blends.
This model displays non-conventional glassy dynamics. 
Mean squared displacements and density-density correlators for the fast component 
exhibit many of the precursor effects characterizing nearby MCT higher-order transitions, 
or fold transitions with exponent parameter very close to unity.

These anomalous features are analogous to those observed for short-ranged attractive colloids, 
for which a higher-order MCT scenario has been derived, suggesting that such features
might be generally present in systems with several competing mechanisms for dynamic arrest.
We suggest bulk-like caging and confinement for the case of polymer blends.
Chain connectivity extends competition between both mechanisms 
over a wide range of blend compositions. The strong resemblance of dynamic correlators 
with those recently reported in fully atomistic simulations of a real polymer blend strongly supports
the dynamic picture here presented. These results open the possibility of new theoretical approaches
for investigating polymer blend dynamics.


\begin{center}
{\bf ACKNOWLEDGEMENTS}
\end{center}

We thank W. G\"{o}tze, E. Zaccarelli, F. Sciortino, and T. Franosch for useful comments and discussions.
Support from the projects NMP3-CT-2004-502235 (SoftComp), 
MAT2004-01017 (Spain), and 206.215-13568/2001 (GV-UPV/EHU Spain) is acknowledged.


\begin{thebibliography}{9}
\bibitem{noteorder} A MCT transition of order $n$, A$_n$, is characterized by $n$ solutions
of the critical non-ergodicity parameter, only the highest value being physical.
\bibitem{notehq} As also found for fold transitions \cite{mctrev1,mctrev2,das},
the coefficients $h_q$ and $h_q^{(2)}$ are in antiphase with $f^{\rm c}_q$.
\bibitem{notemsd} Strictly speaking, sublinear behavior is derived from MCT 
by expanding the mean squared displacement just to first-order.
\bibitem{notefqc} A correction to $f^{\rm c}_q$, of the order of the distance to the transition point,
is neglected in Eq. (\ref{eqlog}).
\end{thebibliography}
\end{document}